\documentclass[prl,aps,twocolumn,showpacs]{revtex4}
\usepackage{epsfig}
\begin{document}
\title{Symmetry breaking and coarsening of clusters in a prototypical driven granular gas}
\author{Eli Livne$^{1}$, Baruch Meerson$^{1}$
and Pavel V. Sasorov$^{2}$} \affiliation{$^{1}$Racah Institute  of
Physics, Hebrew University of  Jerusalem, Jerusalem 91904,
Israel}\affiliation{$^{2}$Institute of Theoretical and
Experimental Physics, Moscow 117259, Russia}
\begin{abstract}
Granular hydrodynamics predicts symmetry-breaking instability in a
two-dimensional (2D) ensemble of nearly elastically colliding
smooth hard spheres driven, at zero gravity, by a rapidly
vibrating sidewall. Super- and subcritical symmetry-breaking
bifurcations of the simple clustered state are identified, and the
supercritical bifurcation curve is computed. The cluster dynamics
proceed as a coarsening process mediated by the gas phase. Far
above the bifurcation point the final steady state, selected by
coarsening, represents a single strongly localized densely packed
2D cluster.

\end{abstract}
 \maketitle
%\section{I. Introduction}
\textit{Introduction.} Rapid granular flow continues to attract a
great deal of attention of physicists \cite{Jaeger}. Among most
fascinating phenomena here is clustering: nucleation and growth of
dense granular clusters, surrounded by dilute granular gas in
``freely-cooling" \cite{Hopkins} and driven
\cite{Kadanoff1,Grossman,Esipov,Kudrolli,Urbach} granular gases.
Clustering can be viewed as a variant of thermal condensation
instability, encountered also in gases and plasmas that cool by
their own radiation \cite{plasma}. This analogy brings about the
question of universality of structure formation in ensemble of
particles with energy losses of different nature. A related,
largely unexplored issue is cluster coarsening. Reviewing
different cluster-forming granular systems with a fixed number of
particles \cite{Hopkins,Urbach,Aranson}, one notices that
coarsening is ubiquitous. In most of these systems a
\textit{single} cluster usually survives after transients die out.
The present work addresses novel issues of pattern formation and
coarsening in cluster-forming driven granular gases. We shall
consider a simple (indeed, prototypical) system: a 2D ensemble of
inelastically colliding smooth hard spheres, confined in a
rectangular box and driven by a rapidly vibrating side wall at
zero gravity. The other three walls are assumed elastic. Though
this and related systems were investigated theoretically
\cite{Kadanoff1,Grossman,Brey,Tobochnik} and experimentally
\cite{Kudrolli,Kudrolli2}, it has been recognized only recently
that they exhibit nontrivial pattern-forming properties
\cite{LMS,KM}. These properties will be in the focus of this work.
We identify the character of symmetry-breaking bifurcations of the
simple clustered state and show that, depending on the control
parameters, both super- and subcritical bifurcations can occur. We
find that selection of the final steady state occurs via cluster
coarsening dynamics, mediated by the gas phase. Far above the
bifurcation point, only one densely packed 2D cluster survives. We
shall conclude that cluster selection by coarsening is universal
in cluster-forming granular flows with a fixed number of
particles.

\textit{Model.} Let $N \gg 1$ identical smooth hard spheres of
diameter $d$ and mass $m=1$ are rolling without friction on a
smooth horizontal surface of a rectangular box with dimensions
$L_x \times L_y$. The number density of grains is $n (x,y,t)$, the
granular temperature is $T(x,y,t)$. For a submonolayer coverage $n
\le n_c = 2/(\sqrt{3} d^2)$, the (hexagonal) close-packing
density. Three of the walls are immobile, and grain collisions
with them are assumed elastic. The fourth wall (located at
$x=L_x$) is rapidly vibrating, $x=L_x + A \cos \omega t$, and
supplies energy to the system. The energy is being lost via
inelastic hard-core grain collisions characterized by a constant
normal restitution coefficient $r$. Our crucial assumption will be
a {\it strong} inequality $1-r^2 \ll 1$. In the quasielastic
limit, and for small Knudsen numbers, the Navier-Stokes granular
hydrodynamics is expected to be reasonably accurate, even for
large granular densities, as long as the granulate is in a
fluidized state \cite{Grossman}. Therefore, the steady states of
the system can be described by the coarse-grained momentum and
energy balance equations
\begin{equation}
p=const\,,\,\,\, \nabla \cdot (\kappa \, \nabla T) = I \,,
\label{energy1}
\end{equation}
where $p$ is the pressure, $\kappa$ is the thermal conductivity
and $I$ is the rate of energy losses. To make the full use of
hydrodynamics, we shall work in the parameter regime when the
energy supply from the vibrating wall can be represented as a
hydrodynamic heat flux. This requires $A \ll l$, where $l$ is the
mean free path of the particles at the vibrating wall. We shall
also assume a double inequality $T^{1/2}/l \ll \omega \ll
T^{1/2}/A$. The left inequality guarantees the absence of
correlations between successive collisions of particles with the
vibrating wall.
%which enables one to average
%the energy flux over the vibration period $2 \pi/\omega$
The right inequality makes the calculation simpler, but it is not
crucial. The resulting energy flux is \cite{LMS,Kumaran}
\begin{equation}
q =\kappa \,
\partial T/\partial x=(2/\pi)^{1/2}\, A^2 \, \omega^2\,n
\,T^{1/2}\,. \label{wall}
\end{equation}
To close the hydrodynamic model, one needs constitutive relations
(CRs) : $p, \kappa$ and $I$ in terms of $n$ and $T$. The CRs have
been derived systematically only in the dilute limit. Reasonably
accurate CRs in the whole range of densities can be obtained by
employing free volume arguments close to the dense-packing limit,
interpolating between the high- and low-density limits and finding
the fitting constants by comparing the results with particle
simulations \cite{Grossman,Luding}. We shall use the CRs suggested
by Grossman et al. \cite{Grossman} because of their relative
simplicity. A special investigation \cite{KM} showed that the
stability diagram  does not change much if one uses instead the
CRs derived by Jenkins and Richman \cite{JR}.

Eqs. (\ref{energy1}) can be rewritten in terms of one variable:
the (scaled) inverse density $z(x,y)=n_c/n(x,y)$
\cite{Grossman,LMS}. We introduce scaled coordinates ${\mathbf r}
/L_x \to {\mathbf r}$ so that the box dimensions become $1 \times
\Delta$, where $\Delta = L_y/L_x$ is the box aspect ratio. We have
\begin{equation}
\nabla \cdot \left( F(z) \nabla z \right) ={\cal L}\, Q(z)\,.
\label{energy2}
\end{equation}
Introducing $\psi=\int_0^{z}F(z^{\prime})\,dz^{\prime}$, we
rewrite it as
\begin{equation}
\nabla^2\psi={\cal L}\,\tilde{Q}(\psi)\, \label{F20}
\end{equation}
The boundary conditions are
\begin{equation}
\left.\frac{\partial\psi}{\partial x}\right|_{x=0}=
\left.\frac{\partial\psi}{\partial y}\right|_{y=0}=
\left.\frac{\partial\psi}{\partial y}\right|_{y=\Delta}=0
\label{F30}
\end{equation}
and (see Ref. \cite{LMS})
\begin{equation}
\left.\frac{\partial\psi}{\partial x}\right|_{x=1}={\cal L}\,
\tilde{H}\left[\psi(1,y)\right]\, \frac{
\int\limits_0^\Delta\!\int\limits_0^1 \tilde{Q}(\psi)\, dxdy}{
\int\limits_0^\Delta \tilde{H}(\psi(1,y))\, dy}\, . \label{F40}
\end{equation}
Here $\tilde{Q}(\psi)=Q\left[z(\psi)\right]$ and
$\tilde{H}(\psi)=H\left[z(\psi)\right]$, while $F,G,H$ and $Q$ are
functions of $z$ only; they are given in Ref. \cite{LMS}. In the
rest of the paper the symbol ~$\tilde{}$~ will be omitted.
Conservation of the total number of particles yields an equation
for the area fraction $f=N/(L_x L_y n_c)$:
\begin{equation}
f=\frac{1}{\Delta}\, \int\limits_0^\Delta\!\int\limits_0^1\frac{dx
dy }{z(\psi)} \, . \label{F60}
\end{equation}
The steady state problem is fully determined by three scaled
parameters: ${\cal L} = (32/3\gamma)\,(L_x/d)^2\,(1-r^2)$ (where
$\gamma \simeq 2.26$), the area fraction $f$ and aspect ratio
$\Delta$. Notice that the steady-state density distributions are
independent of $A$ and $\omega$. This is in contrast to the case
of a non-zero gravity, where the gravity acceleration, combined
with $A \omega^2$, yields a governing parameter.

\textit{Strip state.} The basic state of the system is the ``strip
state": a laterally symmetric cluster located at the wall $x=0$
\cite{Grossman}. The physics of the strip state is simple. Due to
inelastic collisions, the granular temperature goes down with
increasing distance from the driving wall. To maintain the
pressure balance, the granular density should \textit{increase}
with this distance, reaching the maximum at the opposite (elastic)
wall. The strip state is described by the $y$-independent solution
of Eqs. (\ref{F20}) and (\ref{F30}); we shall call it $z=Z(x)$
that corresponds to $\psi=\Psi(x)$. Notice that  Eq. (\ref{F40})
holds automatically in 1D \cite{LMS}. An example of the strip
state is shown in Fig. 2 (bottom left). A similar state was
observed in experiment \cite{Kudrolli}.

\textit{Symmetry-breaking instability.} The 1D strip state gives
way, by a symmetry-breaking bifurcation, to 2D clustered states.
%Numerical solution of Eqs. (\ref{F20})-(\ref{F60}) gave evidence
%for both super- and subcritical bifurcations \cite{LMS}.
The bifurcation point can be found by linearizing Eqs.
(\ref{F20})-(\ref{F60}) around $\psi=\Psi(x)$. In the framework of
\textit{time-dependent} hydrodynamics, this corresponds to
\textit{marginal stability} of the strip state with respect to
small perturbations along the strip. We write
\begin{equation}
\psi(x,y)=\Psi (x) + \varphi_k(x) \exp (iky)+ \mbox{c.c.}
\label{F120}
\end{equation}
and, after linearization, arrive at a linear eigenvalue problem
for $k=k_c(f)$:
\begin{equation}
\varphi_k^{\prime\prime}-{\cal L}Q_{\Psi}\, \varphi_k -
k_c^2\,\varphi_k=0\, , \label{F130}
\end{equation}
\begin{equation}
\varphi_k^{\prime}(0)=0\, , \label{F140}
\end{equation}
\begin{equation}
\varphi_k^{\prime}(1)-{\cal L}\, \frac{ \int\limits_0^1
\tilde{Q}(\Psi(x))\, dx}{\tilde{H}(\Psi(1))}\,
\left.H_{\Psi}\right|_{x=1}\, \varphi_k (1)=0 \,. \label{F150}
\end{equation}
Here and below $(\dots)_\Psi(x)=
\left.\left[F^{-1}\,d(\dots)/dz\right]\right|_{z=Z(x)}$, and
$(\dots)$ stands for any function. Solving the eigenvalue problem
for a given ${\cal L}$, we obtain the marginal stability curve
$k=k_c(f)$ and corresponding eigenfunctions $\varphi_k(x)$. The
modes with $k<k_c(f)$ are unstable. At fixed ${\cal L}$, the
instability occurs for $f_1({\cal L})<f<f_2({\cal L})$, where
$k_c(f_1)=k_c(f_2)=0$ \cite{LMS}. The driving force of the
instability is negative effective lateral compressibility of the
gas \cite{KM}. Figure 1 gives an example of the marginal stability
curve in terms of the minimum aspect ratio $\Delta_c (f) =
\pi/k_c(f)$, at which the instability occurs.

\begin{figure}[ht]
\vspace{-0.1in}
%\vspace{0.8 cm}\epsfxsize=5.5cm\epsffile{fork2vm1.eps}
\epsfig{file=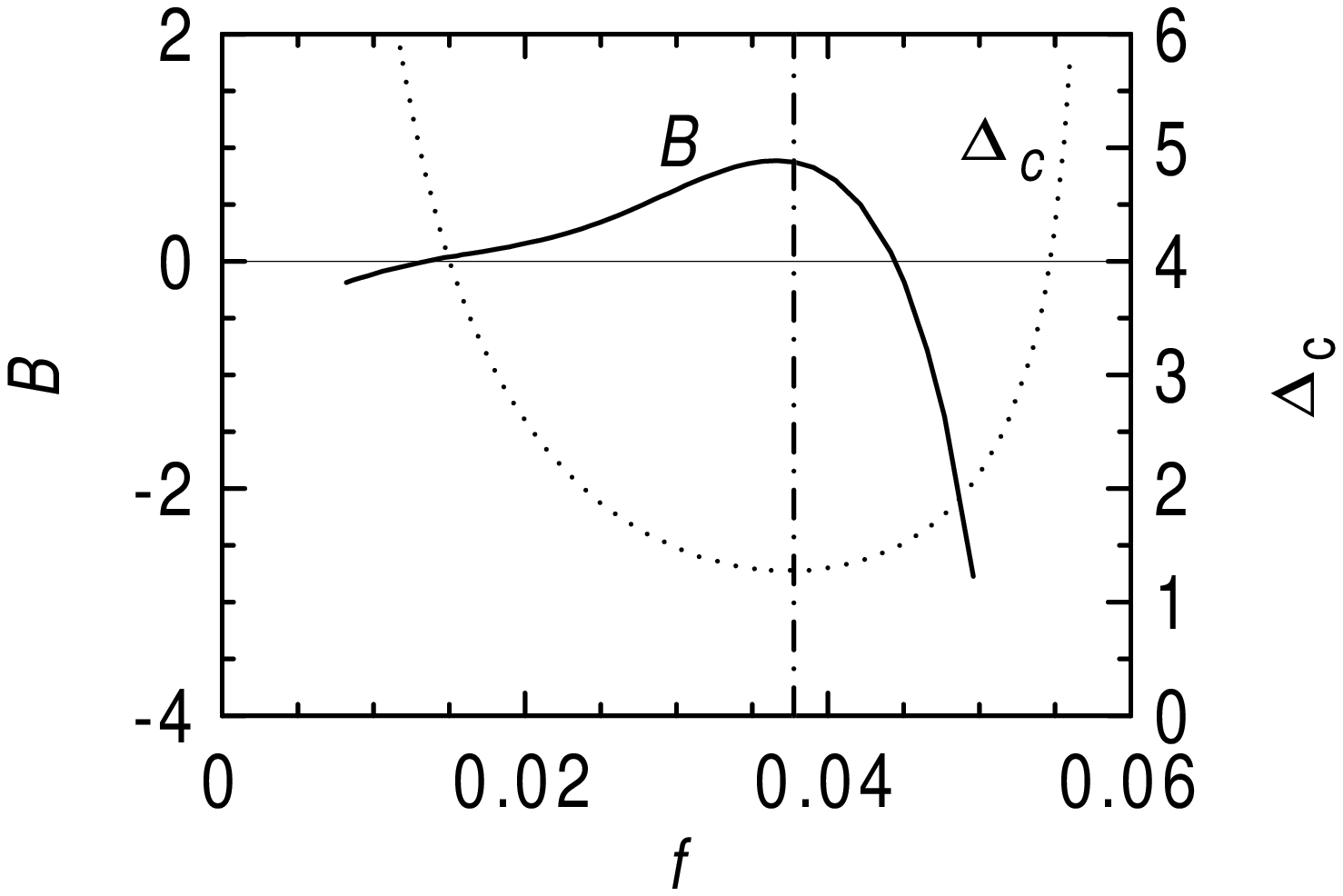, width=5.7cm, clip=} \caption{The
critical aspect ratio $\Delta$ for the instability and parameter
$B$ determining the bifurcation curve (\ref{MT420}) versus the
area fraction $f$ for ${\cal L} =1.25\cdot10^4$. The vertical
dash-dot line corresponds to $f=0.0378$.} \label{Deltac_and_B}
\end{figure}
%\vspace{-0.5cm}
\textit{Bifurcation curve.} To determine the
nature of the bifurcation (sub- or supercritical) and compute the
bifurcation curve, one should go to the second order of the
perturbation theory. We can write
\begin{equation}
\psi(x,y)=\Psi(x) + \sum\limits_n\, \varphi_n(x)\, \exp (inky)\,,
\label{F180}
\end{equation}
where $\varphi_{-n}(x)=\varphi^*_{n}(x)$, and assume that
$\varphi_0\sim\varphi_1^2$, $\varphi_2\sim\varphi_1^2$,
$\varphi_3\sim\varphi_1^3$, etc. Therefore, we need to take into
account only the terms $n=0,\pm 1$ and $\pm 2$. This yields the
following linear equations:
\begin{equation}
\varphi_0^{\prime\prime}-{\cal L}\, Q_{\Psi}\, \varphi_0= {\cal
L}\, Q_{\Psi\Psi}\, \left|\varphi\right|^2 \, , \label{F200}
\end{equation}
$$
\varphi_1^{\prime\prime}-{\cal L}\, Q_{\Psi}\, \varphi_1
-k_c^2\varphi_1 = \left(k^2-k_c^2\right)\, \varphi+
$$
\begin{equation}
+{\cal L}\, \left[Q_{\Psi\Psi}\, \left(\varphi_0\varphi
+\varphi_2\varphi^*\right)+ \frac{1}{2}\, Q_{\Psi\Psi\Psi}\,
\varphi\left|\varphi\right|^2\right]\, , \label{F210}
\end{equation}
\begin{equation}
\varphi_2^{\prime\prime}-{\cal L}\, Q_{\Psi}\, \varphi_2
-4k_c^2\varphi_2 = \frac{1}{2}\, {\cal L}\, Q_{\Psi\Psi}\,
\varphi^2 \, , \label{F220}
\end{equation}
and the boundary conditions:
\begin{equation}
\varphi_0^{\prime}(0)=\varphi_1^{\prime}(0)=\varphi_2^{\prime}(0)
=0\, , \label{F230}
\end{equation}
$$
\varphi_1^{\prime}(1)={\cal L} \Biggl[ \frac{H_\Psi}{H}
\varphi_1\, \int\limits_0^1 Q\, dx \,
$$
$$
+ \frac{\int\limits_0^1 Q\, dx}{H} \left(H_{\Psi\Psi}
\left(\varphi_0\varphi+\varphi_2\varphi^*\right)+
\frac{1}{2}H_{\Psi\Psi\Psi}\, \varphi\left|\varphi\right|^2\right)
$$
$$
+ \frac{H_\Psi}{H} \left(\int\limits_0^1 Q_\Psi\, \varphi_0\, dx+
\int\limits_0^1 Q_{\Psi\Psi}\left|\varphi\right|^2\, dx\right)\,
\varphi
$$
\begin{equation}
- \frac{H_\Psi}{H^2}\, \varphi\left(H_\Psi\, \varphi_0
+H_{\Psi\Psi}\left|\varphi\right|^2\right)\, \int\limits_0^1 Q\,
dx \Biggr]\Biggr|_{x=1}\, , \label{F240}
\end{equation}
\begin{equation}
\varphi_2^{\prime}(1)={\cal L} \left.\left[\frac{H_\Psi}{H}\,
\varphi_2+ \frac{H_{\Psi\Psi}}{2H} \left|\varphi\right|^2
\right]\right|_{x=1}\, \int\limits_0^1 Q\, dx \,, \label{F250}
\end{equation}
where $\varphi=A \,Y(x)$ is a properly normalized solution of Eqs.
(\ref{F130})-(\ref{F150}), see below.  As the boundary condition
for $\varphi_0$ at $x=1$ is fulfilled automatically, one more
condition is needed. For a fixed $f$, this condition is supplied
by Eq. (\ref{F60}):
\begin{equation}
\int\limits_0^1 \frac{\varphi_0}{Z^2F}\, dx=
2\int\limits_0^1\left(\frac{1}{Z^3F^2}+\frac{F_\Psi}{2Z^2F^2}\right)
\left|\varphi_1\right|^2\, dx\, . \label{F260}
\end{equation}
The solvability condition for Eq. (\ref{F210}) yields the
bifurcation curve: a relation between the amplitude of $\varphi_1$
(we call it $A$) and $k_c^2-k^2$. One way to define $A$ is the
following: $\varphi_1 (x) = A \,Y(x)+ A|A|^2\, \delta
\varphi_1(x)$, where $Y(x)$ is the solution of Eqs. (\ref{F130})
and (\ref{F140}) obeying the normalization $Y(0)=1$. This yields
$A\left(k_c^2-k^2\right)=CA|A|^2$,
%\begin{equation}
%A\left(k_c^2-k^2\right)=CA|A|^2\, , \label{F400}
%\end{equation}
where $C=const$. The trivial solution $A=0$ describes the strip
state, while the nontrivial one, $k_c^2-k^2=C|A|^2 $ describes the
bifurcated state. $C>0 \, (<0)$ corresponds to supercritical
(subcritical) bifurcation. The solvability condition is a
generalization of the standard ``orthogonality" condition, or the
Fredholm alternative \cite{Iooss}. It yields $C$ explicitly in
terms of definite integrals of solutions of the homogeneous forms
of Eqs. (\ref{F130}), (\ref{F200}) and (\ref{F220}) that can be
found numerically, see Appendix. We present here the resulting
bifurcation curve for $Y_c$, the (normalized) $y$-coordinate of
the center of mass of the granulate:
\begin{equation}
Y_c= \frac{\int_0^1 dx \int_{-\Delta/2}^{\Delta/2} \, y\, z^{-1}\,
dy }{\Delta \int_0^1 dx \int_{-\Delta/2}^{\Delta/2} z^{-1}\, dy
}\,, \label{cmass}
\end{equation}
where we shifted the $y$-coordinate: $y+\Delta/2 \to y$. %so that
%$y=0$ is now in the midplane.
Let the aspect ratio of the system
$\Delta$ be slightly larger than $\Delta_c=\pi/k_c(f)$ so that
only one mode, with $k=\pi/\Delta$, is unstable. The bifurcation
curve takes the form
\begin{equation}
\left|Y_c\right|= \frac{2}{\pi^2
B^{1/2}}\left(\frac{\Delta}{\Delta_c}-1\right)^{1/2}\, .
\label{MT420}
\end{equation}
where $B=C f^2/(2k_c^2 f_1^2)$ and $f_1=2\int_0^1 Y Z^{-2}F^{-1}
dx$.  Eq. (\ref{MT420}) assumes $B>0$: a supercritical
bifurcation. Figure 1 shows $B(f)$ for ${\cal L}=12,500$. We find
that $B>0$ on an interval of $f$ that lies \textit{within} the
instability interval $(f_1,f_2)$. Closer to the points $f_1$ and
$f_2$ we obtain $B<0$ which indicates subcritical bifurcation.
Subcritical bifurcations close to the high-density instability
border were previously observed by solving numerically the
nonlinear steady state equations (\ref{F20})-(\ref{F40})
\cite{LMS}.

\begin{figure}[ht]
\begin{tabular}{cc}
\hspace{0.5in} \epsfig{file=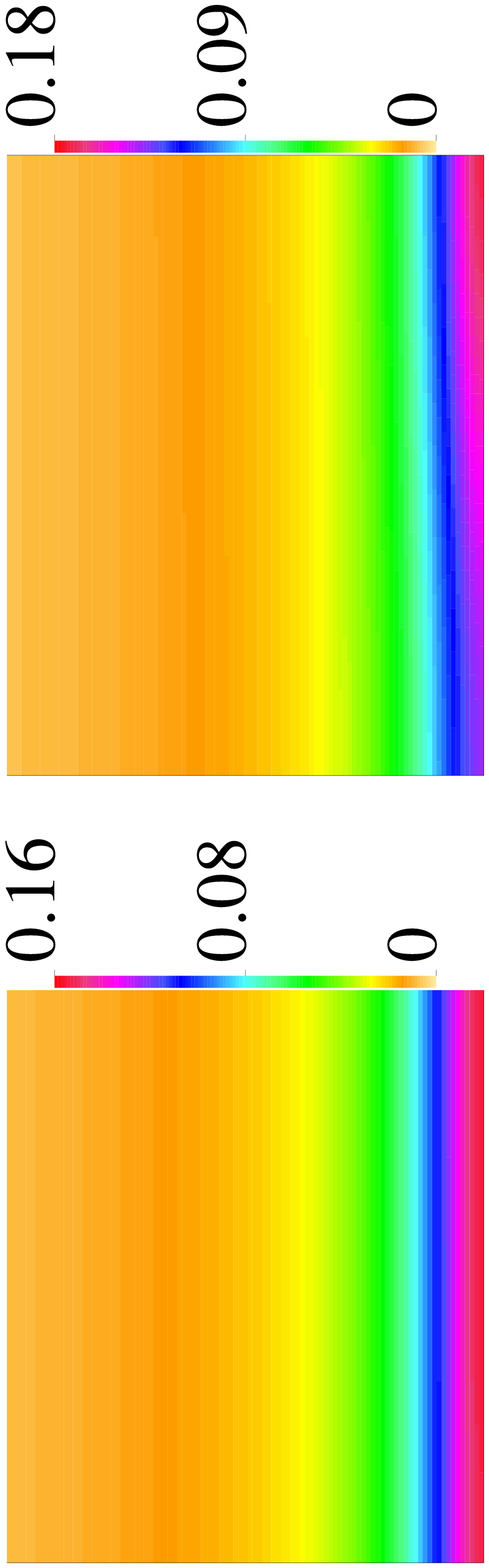, width=1.5cm, clip= } &
\hspace{0.5in} \epsfig{file=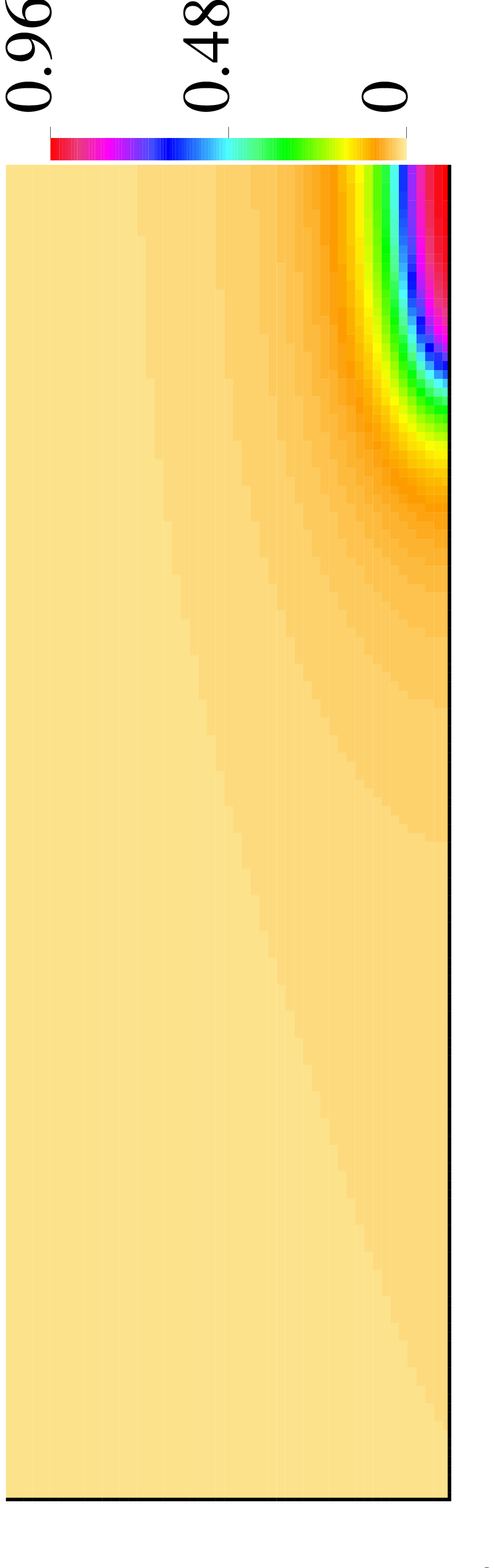, width=1.5cm, clip= }
\end{tabular}
\vspace{0.5cm} \caption{Steady states found in time-dependent
hydrodynamic simulations for ${\cal L}=1.25\cdot10^4\,$,
$f=0.0378$ and $\Delta=1.2$ (bottom left), $1.3$ (top left) and
$3.0$ (right). The left wall of the box is the driving wall.}
\label{fig1}
\end{figure}

\textit{Time-dependent hydrodynamic simulations: bifurcations and
coarsening.} We performed a series of time-dependent hydrodynamic
simulations, to verify the bifurcation theory and to follow the
cluster dynamics at large aspect ratios. The full time-dependent
hydrodynamic equations were solved with the same constitutive
relations and boundary conditions as those used in our steady
state analysis. Instead of the shear viscosity in the
Navier-Stokes equation we accounted for a small model friction
force $- n {\mathbf v}/{\tau}$, where ${\mathbf v}$ is the
hydrodynamic velocity. An extended version of the compressible
hydro code VULCAN \cite{vulcan} was employed.

\begin{figure}[ht]
%\vspace{0.7 cm}
%\epsfxsize=4.5cm  \epsffile{comp.eps}
\epsfig{file=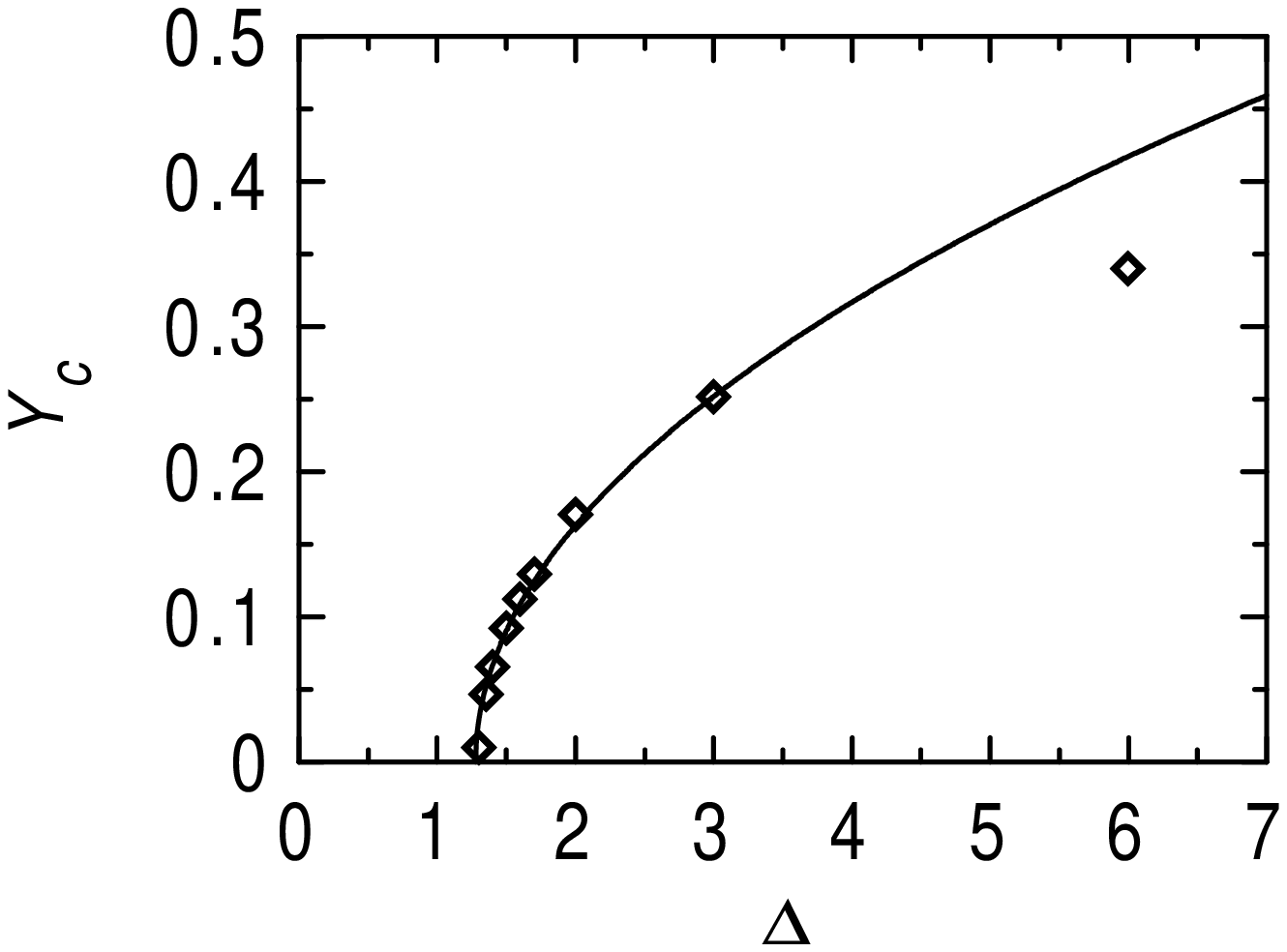, width=4.5cm, clip= } \caption{Bifurcation
curve $Y_c(\Delta)$ predicted by Eq. (\ref{MT420}) (line) and
found in hydrodynamic simulations (squares) for ${\cal L}
=1.25\cdot10^4$ and $f=0.0378$.} \label{bifurcurve}
\end{figure}

We fixed ${\cal L}=1.25\cdot10^4$ and $f=0.0378$ and varied
$\Delta$. The initial scaled density included a zero mode
corresponding to the
fixed $f$ plus small-amplitude random noise. %The initial velocity
%was zero, the initial pressure uniform.
Figure 2 shows the final states for different aspect ratios
$\Delta$. For ${\cal L}$ and $f$ used, the marginal stability
theory predicts $\Delta_c \simeq 1.28$ (see Fig. 1). Indeed, the
strip state observed at $\Delta=1.2$ (Fig. 2, bottom left) gives
way to a slightly asymmetric state at $\Delta=1.3$ (Fig. 2, top
left). Far above the bifurcation point the final state represents
a densely packed 2D cluster (``island"), located in a corner (Fig.
2, right). This implies that all but one of the multiple 2D steady
state solutions found earlier (chains of islands periodic in the
$y$-direction) \cite{LMS} are unstable. The stable steady state
selected by the coarsening dynamics is the one with the maximum
possible period, equal to twice the lateral dimension of the
system.

Figure 3 shows the bifurcation curve $Y_c (\Delta)$ predicted by
Eq. (\ref{MT420}), and measured in the simulations after
transients die out. Excellent agreement is obtained for not too
large supercriticalities.  Close to the bifurcation point we
observed exponential slowdown as expected.

\begin{figure}[ht]
%\vspace{0.5 cm}
%\hspace{1.0cm}
%\epsfxsize=6.5 cm \epsffile{6a1.eps} \vspace{0.2cm}
\epsfig{file=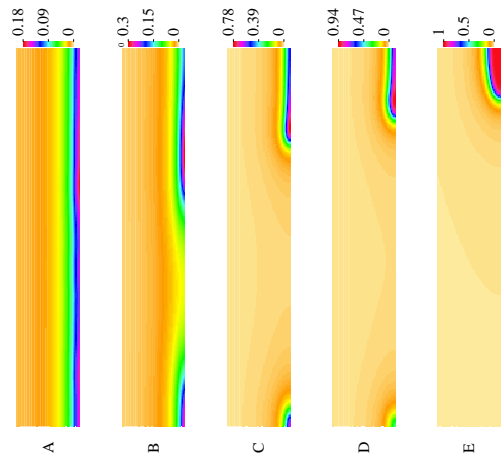, width=6.5cm, clip= } \caption{Time history
of the density field for ${\cal L} =1.25\cdot10^4$, $f=0.0378$ and
$\Delta=5$ at scaled times $650$ (A),  $2,100$ (B), $2,850$ (C),
$3,250$ (D), and $7,000$ (E). The left wall is the driving wall.
Notice the change of color code with time.} \label{coarsening}
\end{figure}

Now we present the simulation results on the cluster dynamics and
selection. For large $\Delta$ the dynamics involve two stages (see
Fig. 4). During the first stage, several clusters nucleate at the
wall opposite to the driving wall (Fig. 4A). Their number is of
the order of $\Delta/\Delta_c$ which apparently corresponds to the
maximum linear growth rate of the instability versus $k$. At the
slower second stage the clusters become denser. As they compete
for the material, their number goes down, and only one densely
packed cluster (``island") finally survives, always in a corner
(Fig. 4e). The clusters interact mostly through the gas phase,
similarly to Ostwald ripening in phase ordering systems with
conserved order parameter, controlled by gasdynamics
\cite{plasma,AMS}. We also observed, for a different realization
of noise in the initial conditions, direct coalescence of
transient clusters.  However, the resulting single cluster is
always the same in simulations with the same ${\cal L},\,f$ and
$\Delta$.

\textit{In summary}, we used hydrodynamics to determine the
character of symmetry-breaking bifurcations in a prototypical
driven granular gas and compute the supercritical bifurcation
curve. We found the selected steady state with broken
translational symmetry and showed that selection is made via a
coarsening process similar to Ostwald ripening. It appears that
cluster selection by coarsening is a universal selection mechanism
in cluster-forming granular flows with a fixed number of
particles.

This research was supported by the Israel Science Foundation,
founded by the Israel Academy of Sciences and Humanities, and by
the Russian Foundation for Basic Research (grant No. 02-01-00734).

\textit{Appendix. Solvability condition}. Here we present the
solvability condition for the boundary value problem described by
the linear equation (\ref{F210}) and boundary conditions
(\ref{F230}) and (\ref{F240}). This solvability conditions yields
the constant $C$ that enters the bifurcation equation
$k_c^2-k^2=C\,|A|^2$. Consider the following problem:
\begin{eqnarray}
w^{\prime\prime}(x)+P(x)w(x)& = &f(x)\, ,
\label{F270}\\
w^{\prime}(0)& =& 0\,, \label{F290} \\
w^{\prime}(1)+aw(1)& =& g\,
, \label{F280}
\end{eqnarray}
where $g$ is a parameter; and assume that the homogeneous variant
of this problem,
\begin{eqnarray}
w^{\prime\prime}(x)+P(x)w(x)& = & 0\, ,
\label{F300}\\
w^{\prime}(0)& =& 0 \, ,\label{F320}\\
w^{\prime}(1)+aw(1)& =& 0
\label{F310}
\end{eqnarray}
has a nontrivial solution. The problem~(\ref{F270})-(\ref{F280})
will have no solution unless there is a relation between $f(x)$
and $g$. What is the relation? To answer this question, we
construct the solution of the problem~(\ref{F270})-(\ref{F280})
using two fundamental solutions $w_0$ and $w_1$ of the homogeneous
equation~(\ref{F300}). We choose these solutions as follows.  Let
$w_0$ obeys the boundary conditions ~(\ref{F320}) and
\begin{equation}
w_0(0)=1\, .\label{F330}
\end{equation}
In view of the above assumption, Eq.~(\ref{F310}) is fulfilled
automatically. The second fundamental solution $w_1$ of
Eq.~(\ref{F300}) obeys the following boundary conditions (again at
$x=0$):
\begin{equation}
w_1(0)=0\,, \quad w_1^{\prime}(0)=1\,. \label{F340}
\end{equation}
Notice that each of the solutions $w_0$ and $w_1$ is defined by an
initial value problem, rather than by a boundary value problems.

The general solution of Eqs.~(\ref{F270})-(\ref{F280}) can be
written as $C_0(x)\,w_0(x) + C_1(x)\,w_1(x)$, where functions
$C_0(x)$ and $C_1(x)$ are yet unknown, and we can demand
$C_0^{\prime} (x) \,w_0(x) + C_1^{\prime} (x) \,w_1(x)=0$.
Straightforward calculations yield the solvability condition
\begin{equation}
g=\left[w_1^{\prime}(1) + aw_1(1)\right]\int\limits_0^1 fw_0\,
dx\, . \label{F350}
\end{equation}
In the particular case of $g=0$ Eq. (\ref{F350}) is reduced to the
standard orthogonality condition  $\int\limits_0^1 fw_0\, dx\,=0$
\cite{Iooss}.

Applying the solvability condition (\ref{F350}) to
Eq.~(\ref{F210}) with the boundary conditions~(\ref{F230}) and
(\ref{F240}), we obtain the following relationship:
$$
{\cal L} \Biggl\{\left(\frac{H_{\Psi\Psi}}{H}
(\varphi_0\varphi+\varphi_2\varphi^*)
+\frac{H_{\Psi\Psi\Psi}}{2H}\,
\varphi\left|\varphi\right|^2\right) \int\limits_0^1 Q\, dx+
$$
$$
+\frac{H_{\Psi}}{H}\varphi \left(\int\limits_0^1 Q_{\Psi}\,
\varphi_0\, dx+ \int\limits_0^1
Q_{\Psi\Psi}\left|\varphi\right|^2\, dx\right)-
$$
$$
-\left(\frac{H_{\Psi}^2}{H^2}\varphi\varphi_0+
\frac{H_{\Psi}H_{\Psi\Psi}}{H^2}
\varphi\left|\varphi\right|^2\right) \int\limits_0^1 Q\,
dx\Biggr\}\Biggr|_{x=1}=
$$
$$
=\left.\left[Y_{11}^{\prime}-{\cal L}\frac{H_{\Psi}}{H}Y_{11}
\int\limits_0^1 Q\, dx \right]\right|_{x=1} \int\limits_0^1
Y_{10}\biggl[\left(k^2-k_c^2\right)\varphi +
$$
\begin{equation}
+{\cal
L}\left(Q_{\Psi\Psi}\left(\varphi_0\varphi+\varphi_2\varphi^*\right)
+\frac{1}{2}\, Q_{\Psi\Psi\Psi}\,
\varphi\left|\varphi\right|^2\right) \biggl]\, dx\, , \label{F360}
\end{equation}
where $Y_{0}(x)$ is solution of Eq.~(\ref{F130}) with the boundary
conditions
\begin{equation}
Y_{0}(0)=1\, ,\quad Y_{0}^{\prime}(0)=0\, , \label{F370}
\end{equation}
and $Y_{1}(x)$ is the solution of Eq.~(\ref{F130}) with the
boundary conditions:
\begin{equation}
Y_{1}(0)=0\, ,\quad Y_{1}^{\prime}(0)=1\, . \label{F380}
\end{equation}
Notice that each of the solutions $Y_{0}(x)$ and $Y_1(x)$ is
defined by an initial value problem, rather than by a boundary
value problems. This makes the numerical solution straightforward.
To complete the calculation, we solve numerically the
initial-value problems for the linear differential equations for
$\varphi$, $\varphi_0$ and $\varphi_2$ and compute the definite
integrals entering Eqs. (\ref{F260}) and (\ref{F360}). The final
result is given by Eq. (\ref{MT420}).


\begin{references}
\bibitem{Jaeger} H.M. Jaeger, S.R. Nagel, and R.P. Behringer, Rev.
Mod. Phys. {\textbf 68}, 1259 (1996); L.P. Kadanoff, Rev. Mod.
Phys. {\textbf 71}, 435 (1999).
\bibitem{Hopkins} M.A. Hopkins and M.Y. Louge, Phys. Fluids A {\textbf 3}, 47 (1991); I. Goldhirsch
and G. Zanetti, Phys. Rev. Lett. 70, 1619 (1993); S. McNamara and
W.R. Young, Phys. Rev. E {\textbf 53}, 5089 (1996).
\bibitem{Kadanoff1} Y. Du, H. Li, and L.P. Kadanoff, Phys. Rev.
Lett.  {\textbf 74}, 1268 (1995).
\bibitem{Grossman} E.L. Grossman, T. Zhou, and E. Ben-Naim, Phys. Rev. E {\textbf 55},
4200 (1997).
\bibitem{Esipov} S.E. Esipov and T. P\"{o}schel, J. Stat. Phys. {\textbf 86}, 1385 (1997).
\bibitem{Kudrolli} A. Kudrolli, M. Wolpert, and J.P. Gollub, Phys. Rev. Lett. {\textbf 78}, 1383 (1997).
\bibitem{Urbach} J.S. Olafsen and J.S. Urbach, Phys. Rev. Lett. \textbf{81}, 4369 (1998).
\bibitem{plasma} B. Meerson, Rev. Mod. Phys. {\textbf 68}, 215 (1996).
\bibitem{Aranson} I. S. Aranson \textit{et al.}, Phys. Rev. Lett. {\textbf 84}, 3306
(2000). %I. Aranson  \textit{et al.}, e-print cond-mat/0107443.
\bibitem{Brey} J.J. Brey and D. Cubero, Phys. Rev. E {\textbf 57}, 2019 (1998).
\bibitem{Tobochnik} J. Tobochnik, Phys. Rev. E {\textbf 60}, 7137 (1999).
\bibitem{Kudrolli2}A. Kudrolli and J. Henry, Phys. Rev. E {\textbf 62}, R1489 (2000).
\bibitem{LMS} E. Livne, B. Meerson, and P.V. Sasorov, Phys. Rev. E {\textbf 65}, 021302 (2002).
\bibitem{KM} E. Khain and B. Meerson, cond-mat/0201569.
%\bibitem{opposite}  In the opposite limit, $A \, \omega \gg
%T^{1/2}$, one obtains $q = 2 \pi A \omega n T$. An
%interpolation formula $q= 2\pi A \omega \,n T\,(1+2^{1/2}
%\pi^{3/2} \epsilon)^{-1}$ should work well for any $\epsilon=
%T^{1/2}/(A \omega)$.
\bibitem{Kumaran} V. Kumaran, Phys. Rev. E {\textbf 57},
5660 (1998).
\bibitem{Luding} S. Luding, Phys. Rev. E \textbf{63}, 042201 (2001).
\bibitem{JR} J.T. Jenkins and M.W. Richman, Phys. Fluids {\textbf 28}, 3485 (1985).
%\bibitem{dilute} J.J. Brey and D. Cubero, in \textit{Granular Gases}, edited by
%T. P\"{o}schel and S. Luding (Springer, Berlin, 2001), pp. 59-78;
%I. Goldhirsch, \textit{ibid}, pp. 79-99.
\bibitem{Iooss} G. Iooss and D.D. Joseph, \textit{Elementary Stability
and Bifurcation Theory} (Springer, New York, 1980), p. 88.
\bibitem{vulcan} E. Livne, Astrophys. J. {\textbf 412}, 634 (1993).
%\bibitem{Bray} A.J. Bray,  Adv. Phys. {\textbf 43}, 357 (1994).
\bibitem{AMS} I. Aranson, B. Meerson and P.V. Sasorov, Phys. Rev. E \textbf{52}, 948 (1995).

\end{references}
\end{document}